# SIMULATION OF A TYPICAL HOUSE IN THE REGION OF ANTANANARIVO, MADAGASCAR DETERMINATION OF PASSIVE SOLUTIONS USING LOCAL MATERIALS

Harimalala Razanamanampisoa*, Zely Arivelo Randriamanantany, Hery Tiana Rakotondramiarana
Institut pour la Maîtrise de l'Energie Université d'Antananarivo
*harimanampy@yahoo.fr

François Garde, Harry Boyer
Laboratoire de Physique des Bâtiments et des Systèmes Campus Sud -Université de La Réunion

ABSTRACT

This paper deals with new proposals for the design of passive solutions adapted to the climate of the highlands of Madagascar. While the strongest population density is located in the central highlands, the problem of thermal comfort in buildings occurs mainly during winter time. Currently, people use raw wood to warm the poorly designed houses. This leads to a large scale deforestation of the areas and causes erosion and environmental problems. The methodology used consisted of the identification of a typical building and of a typical meteorological year. Simulations were carried out using a thermal and airflow software (CODYRUN) to improve each building component (roof, walls, windows, and soil) in such a way as to estimate the influence of some technical solutions on each component in terms of thermal comfort. The proposed solutions also took into account the use of local materials and the standard of living of the country.

Keywords: Madagascar, Insulation, Passive solutions, Thermal comfort

1. INTRODUCTION

The island of Madagascar is located in the Indian Ocean, Austral hemisphere. It stretches over a length of 1,650 km from latitude 12° to 25° South. Madagascar offers a broad diversity of climates ranging from humid cold to dry heat through humid heat [1].

Our study concerns the center of the island, more precisely, the region of Analamanga, which is part of the temperate altitude climatic regime, superior to 1,200 meters. This climatic domain encompasses the central axis of the highlands and covers a large part of the Province of Antananarivo. The year is divided into two well individualized seasons: a rainy and averagely warm season from November to March, and a cool and relatively dry season during the rest of year. Therefore the problem of thermal comfort in buildings occurs mainly during this winter period.

From a demographic point of view, the national density of the population is 22 inhab/km$^2$. Moreover, Antananarivo, the capital city, is home to the largest number of the people of the island and the global population density in this region is in the range of 93 inhab/km$^2$, which is almost 4 times superior to the national average [2]. However, the majority of the population lives in poorly designed houses.

In order to face these problems, new houses should be carefully designed and built to ensure thermal comfort without using active solutions. The aim of this work is to offer simple technical passive solutions at an acceptable cost which also take into account the use of local materials and the living standard of the country. Computer simulations were used to study the thermal behavior of a typical house with the use of a code building dynamic thermal CODYRUN [3].

2. METHODOLOGY

The following paper illustrates the adopted methodology:
- Research of a typical meteorological year in the region of Antananarivo;
- Research of a typical house representative of what is nowadays constructed as pilot project ;
- Simulations with the typical meteorological file by making the constituting parameters of the typical house vary, and particularly monitoring the daily evolution of the resultant temperature in the dwelling.
- Exploitation and analysis of the results by zone.

Various authors have worked on specific problems concerning the materials and components used for the construction of buildings; and these studies were carried out in some countries having a climate similar to that of the highlands of Madagascar, namely South Africa [4] and China [5].

## 2.1 Definition of the typical sequences

The study of the effects of external conditions on a building requires the availability of hourly meteorological data, representative of the studied site. The typical meteorological year represents the climatic solicitations of the studied site. It is from this typical meteorological year that the climatic winter sequence was selected because it is the most representative of the conditions of the cold season as illustrated in Fig. 1. In fact, it is in this cool period of the year that the most unfavorable combination of climatic parameters was found: extremely cold sequence (absolute minimal temperature of 5.6°C) accompanied with humidity at night, and the alternation of periods of sunshine during the day.

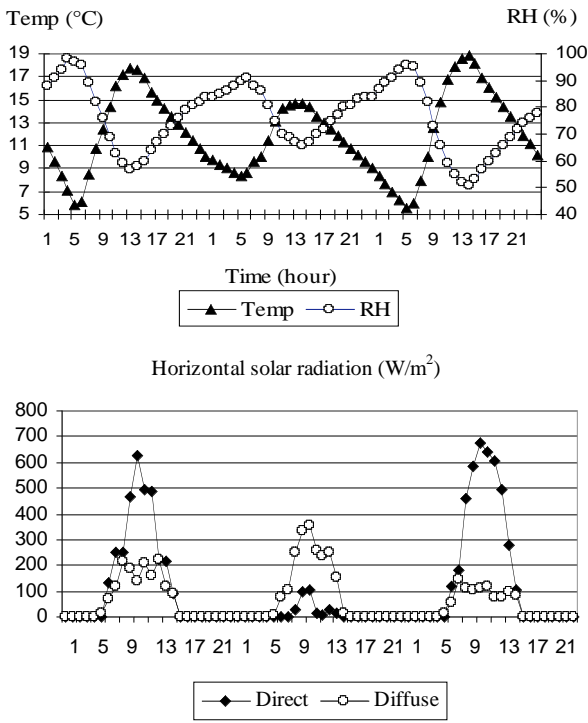

Fig. 1 Typical sequences in the simulations

The characteristic values of the climatic agents for the site of Antananarivo are:

Number of hours of sunshine : 2583 hours;
Average annual temperature : 19.3-19.8 °C
Average maximum temperature – Hot season: 25.9°C
Average minimum temperature – Hot season: 15.9°C
Average maximum temperature – Cold season: 20.6°C
Average minimum temperature – Cold season: 9.6°C

## 2.2 Choice of the typical house

The chosen typical house is that which is most representative of the type of dwelling built in Antananarivo, in terms of architecture and number of rooms. The house is composed of two bedrooms, a living-room, a toilet, a bathroom and a kitchen. Its plan is represented in Fig. 2 and its materials in Table 1. As for basic orientation, the living-room faces north. The surfaces of each room are as follows: bedroom 1, east façade: 7.308m$^2$, glazing: 1.26m$^2$ and south façade: 12.58m$^2$, bedroom 2, south façade: 12.58m$^2$, glazing: 1.26m$^2$ and living-room: west façade: 10.43m$^2$ and north façade: 19.12m$^2$, glazing :1.26 m$^2$.

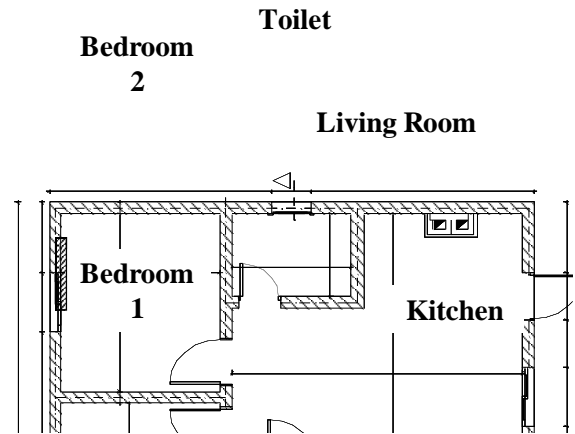

Fig. 2: Plan of typical house

**TABLE 1: MATERIALS OF THE TYPICAL HOUSE**

| Components | Material | Thickness (m) | Lambda (W.m$^{-1}$. °C$^{-1}$) |
|---|---|---|---|
| Walls | Bricks not burned | 0.22 | 0.69 |
| Doors | Pine | 0.035 | 0.16 |
| Roof | Tile | 0.015 | 0.60 |
|  | Air | 0.17 | 0.85 |
|  | Plaster | 0.005 | 0.29 |

## 2.3. Comparison criterion

In order to compare the different technical solutions, the chosen thermal comfort parameter is the resultant temperature. This variable allows taking into account the non-comfort arising from the long wave radiative effects [6]. The dynamic evolution of this variable throughout the typical sequence was monitored, but average day and night temperatures were also used, characterizing the use of day rooms (the living room) and the use of night rooms (the bedrooms). Resultant day temperature: average resultant temperature from 07:00 h to 18:00 h (Tres day); resultant night temperature: average resultant temperature from 19:00 h to 06:00 h (Tres night).

3. RESULTS AND DISCUSSION

As far as outer protecting structures are concerned (roof, opaque walls, glazing, soil), an initial phase of the simulations was made to see the thermal behavior of the basic typical house by means of the resultant temperature without using the passive solutions. During simulation, the typical house is closed and the only system of air renewal is composed of small openings: orifices (holes, joints) separating two thermal zones or separating a zone and the outside of the building. This first result shows that the resultant temperature, which defines the thermal comfort, is largely below the accepted limit. Fig. 3 summarizes the total of non-comfort hours in each zone.

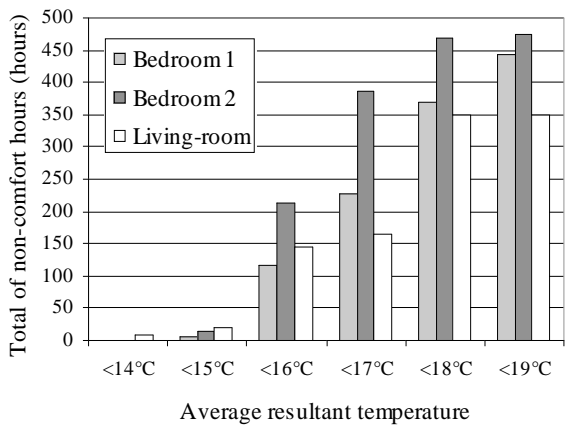

Fig. 3: Total of non-comfort hours of the basic typical house during winter time

3.1 Orientation and surface of glazing - soil

In order to improve the conditions of winter comfort, it is necessary to optimize the solar gains to benefit from free solar supply by varying the orientation and the glazing [7]. It is necessary to see the impact of all orientations and surfaces of glazing on the thermal behavior of the building.

The results show that the resultant temperature 17°C for rooms reached a livable level for the night. As for the living-room, 19 / 20°C is a good result for the day.

Color plays an important role in the thermal behavior of a building due to its absorption of the solar radiation. In order to see the impact of the color of the soil on the resultant temperature, simulations were made on the typical house oriented southward, with glazing percentages of 30 % (N_ E); 10 % S; 20 % W and two different absorption coefficients (0.75, 0.9) of the soil. The percentage is that of the surface of glazing with regard to the surface of the façade. The thermal behavior of all the zones has improved by 1°C to 1.5°C.

**TABLE 2: EXPLANATION OF THE LEGEND OF THE FIGURES**

| Numbering | Surface of glazing |
|---|---|
| 0 | The basic typical house |
| 1 | 10% (N_E_S_W) |
| 2 | 20% (N_E_S_W) |
| 3 | 30% (N_E_S_W) |
| 4 | 30% (N_E) ; 20% S ; 30% W |
| 5 | 30% (N_E) ; 10% S ; 30% W |
| 6 | 30% (N_E) ; 10% S ; 20% W |

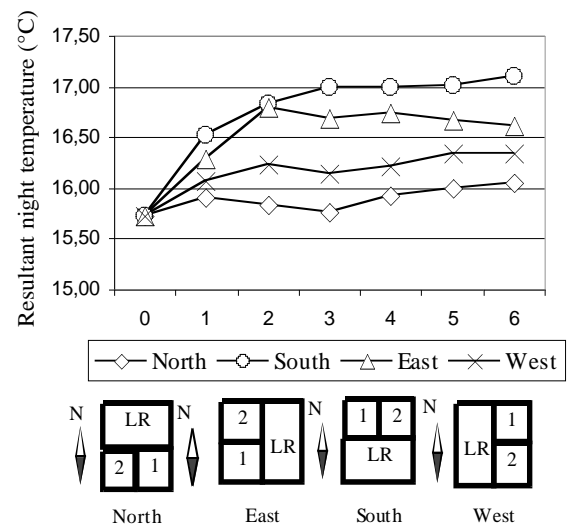

Fig. 4: The effect of orientation and surface of glazing on the resultant night temperature of the bedroom 2

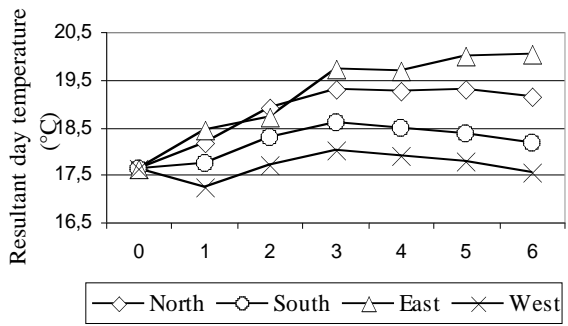

Fig. 5: The effect of orientation and surface of glazing on the resultant day temperature of the living-room

3.2 Insulation of roof and walls

For the zones of temperate altitude climate, the objective is to reduce the fluctuations of the temperature of internal air and to improve the conditions of comfort at night. Insulation is strongly recommended for this type of climate, in order to reduce these fluctuations. The insulating materials used are the local materials found in the capital city. Straw was used for the roof and a mixture of earth or clay coating with the straw, also known as "torchi", for the walls. Fig. 6-7 summarize the evolution of the resultant night temperature (bedrooms 1 and 2) and the resultant day temperature (living -room) according to the thickness of the straw and torchi. As a whole, results show that when the thickness of straw is 15 cm and above, the increase of the resultant temperatures is quasi-constant.

With only 15 cm of straw or torchi, the insulation of the roof allows to make the resultant temperature increase of 1.5°C for the bedrooms and of 1°C for the living-room, whereas for the insulation of walls, it increases the resultant temperature by 1.5°C for the bedrooms and 2°C for the living-room.

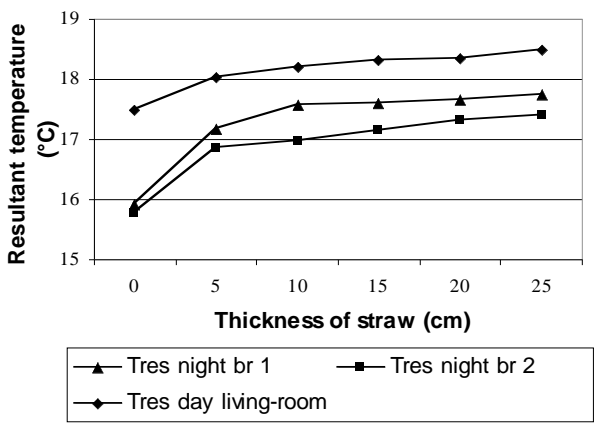

Fig. 6: Insulation of roof for the bedrooms and the living-room

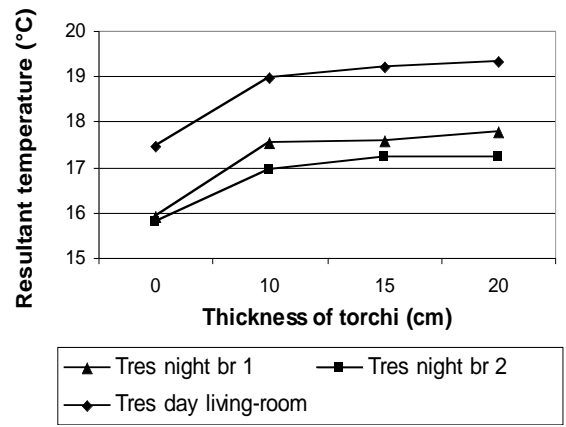

Fig. 7: Insulation of walls for the bedrooms and the living-room

4. CONCLUSION

The purpose of this article has been to offer passive simple technical solutions using local materials and at a low cost in the case of the region of Antananarivo Madagascar, which is located in a temperate altitude climate. All performed simulations led to define competitive passive technical solutions for every component of the envelope: straw for the insulation of the roof, torchi for the internal and external insulation of the walls. All the orientations coupled with the different percentages of surface of glazing have been visualized, as well as the importance of the color of the soil. With the simple passive solutions, i.e. 15 cm of straw, 15 cm of torchi, the optimal orientation towards the South for the case of the typical house, and the surface of glazing with regard to façades of 30% (N_E); 10% S; 20% W, the comfort is reached 100 % of the time both at night and during the day. The proposition of the application of the local passive solutions as well as of the local know-how meets the general policy of the State because not only does it focus on environmental protection, but it also contributes to setting construction standards. Future research should be carried out in the field of passive solutions for the benefit of the other provinces of Madagascar.

5. ACKNOWLEDGEMENTS

Heartfelt thanks go to the AUF-EDIM Madagascar Project for their financial contribution to this study.